\documentclass[twocolumn,a4paper,showpacs,preprintnumbers,amsmath,amssymb,nofootinbib,floatfix]{revtex4}
\usepackage{epsfig}
\input psfig.sty

\usepackage{graphicx}
\usepackage{dcolumn}
\usepackage{bm}
\usepackage{epsfig}
\usepackage{graphicx}
\usepackage{dcolumn}
\usepackage{bm}
\usepackage[english]{babel}

\newcommand{\mytilde}{\raise.17ex\hbox{$\scriptstyle\mathtt{\sim}$}}

\newcommand{\aap}{Astron. and Astrophys.}

\newcommand{\no}{\mbox{$n_{e_o}$}}

\newcommand{\Xo}{\mbox{$S_{x  0}$}}

\newcommand{\LameH}{\mbox{$\Lambda_{e \mbox{\tiny H}}$}}

\newcommand{\Da}{\mbox{$D_{\!\mbox{\tiny A}}$}}

\begin{document}
\title{Is the non-isothermal double $\beta$-model incompatible with no time evolution of galaxy cluster gas mass fraction ?}
\author{R. F. L. Holanda$^{1,2,3}$  } \email{holandarfl@gmail.com}
\affiliation{$^1$Departamento de F\'{\i}sica, Universidade Federal de Sergipe, 49100-000, Sao Cristovao - SE, Brasil
\\$^2$ Departamento de F\'{\i}sica, Universidade Federal de Campina Grande, 58429-900, Campina Grande - PB, Brasil
\\$^3$ Departamento de F\'{\i}sica, Universidade Federal do Rio Grande do Norte, 59078-970, Natal - RN, Brasil}

\date{\today}

\begin{abstract}

In this paper, we propose a new method to obtain the depletion factor $\gamma(z)$, the ratio by which the measured baryon fraction  in galaxy clusters is depleted with respect to the universal mean. We use exclusively  galaxy cluster data, namely, X-ray gas mass fraction ($f_{gas}$) and angular diameter distance measurements from Sunyaev-Zel'dovich effect plus X-ray observations. The galaxy clusters are the same in both data set and the non-isothermal spherical double $\beta$-model was used to describe their electron density and temperature profiles. In order to compare our results with those from recent cosmological hydrodynamical simulations, we suppose a possible time evolution for $\gamma(z)$, such as, $\gamma(z) = \gamma_0(1 + \gamma_1z)$. As main conclusions we found that: the $\gamma_0$ value is in full agreement with the simulations. {  On the other hand, although  the $\gamma_1$ value found in our analysis is compatible with $\gamma_1=0$ within 2$\sigma$ c.l., our results show a non-negligible time evolution for the depletion factor, unlike the results of the simulations. However, we also put constraints on $\gamma (z)$ by using the $f_{gas}$ measurements and angular diameter distances  obtained from the flat $\Lambda$CDM model (Planck results) and from a sample of galaxy clusters described by a elliptical profile. For these cases no significant time evolution for $\gamma(z)$ was found. Then, if a constant depletion factor  is an inherent characteristic of these structures, our results show that   the spherical double $\beta$-model used to describe the galaxy clusters considered  does not affect the quality of their $f_{gas}$ measurements.}
  
\end{abstract}
\maketitle

\section{Introduction}

The galaxy clusters are the largest gravitationally bound
structures in the Universe and their observations are powerful
tools to probe the evolution of the Universe at redshifts $z < 2$.
For instance, the evolution of their X-Ray temperature
and luminosity functions can be used
to limit the matter density, $\Omega_M$, and the normalization of
the density fluctuation power spectrum, $\sigma_8$ (Henry 2000; Ikebe et al. 2002; Mantz et a. 2014).
It is expected that the abundance of galaxy clusters as a function of mass and
redshift imposes restrictive limits on dark energy
models (Albrecht et al. 2006; Basilakos, Plionis \& Lima
2010; Chandrachani Devi \& Alcaniz, 2014). Multiple
image systems behind galaxy clusters can also be used to
estimate cosmological parameters via strong gravitational
lensing (Lubini et al. 2013). The combination of the
X-ray emission of the intra-cluster medium with the Sunyaev-
Zel'dovich effect (Sunyaev-Zel’dovich 1972) provides  angular diameter distances, $D_A$, of galaxy clusters (Reese et al. 2002; De Filippis et al. 2005; Bonamente et al. 2006), and, consequently, a Hubble diagram {  (see details in Allen, Evrard \& Mantz 2011; Huterer \& Shafer 2017; Czerny et al. 2018).}

Particularly, {  if one assumes that the X-ray gas mass
fraction, $f_{gas}$, of hot, massive and relaxed galaxy clusters does not evolve with redshift, it  can be used to constrain cosmological
parameters (Sasaki 1996; Pen 1997; Ettori et al. 2003, 2004, 2006, 2009; Allen et al. 2002, 2004, 2008; Gon\c{c}alves et al. 2012; Mantz et al. 2008; Mantz et al. 2014). This assumption rises due to galaxy clusters are relatively well isolated systems and the largest gravitationally bound structures in the Universe. Moreover, in the absence of dissipation, the ratio of baryonic-to-total in clusters should closely match the ratio of the cosmological parameters measured from the cosmic microwave background. So, in this kind of test, knowing the gas content in galaxy clusters and whether it  evolves with redshift  is a key ingredient.  At this point, it is worth to comment that Lagana et al. (2013) investigated the baryon distribution in groups and clusters, more precisely,  123 systems  in redshift range $0.02 < z < 1$. They found that the gas mass fraction does not depend on the total mass  for systems  more massive than $10^{14}$ solar masses. Moreover,  only a slight dependence of gas mass fraction measurements with redshift  for $r_{2500}$\footnote{This radii is that one within which the mean cluster density is 2500 times the critical density of the Universe at the cluster's redshift.} was obtained (see their fig. 6).
 
In order to quantify the gas content and its possible time evolution, hydrodynamic simulations have been used to calibrate the baryon depletion factor, $\gamma(z)$, the ratio by which the baryon fraction of galaxy clusters is depleted with respect to the  universal  mean of baryon fraction\footnote{These hydrodynamic simulations consider the flat $\Lambda$CDM, with $\Omega_M=0.24$ for the matter density parameter, $\Omega_b = 0.04$ for the contribution of baryons, $H_0 = 72$ km/s/Mpc for the present-day Hubble constant, $n_s = 0.96$ for the primordial spectral index and $\sigma_8 = 0.8$ for the normalization of the power spectrum.} (see Evrard 1997; Metzler and Evrard 1994; Bialek et al. 2001; Muanwong et al. 2002; Ettori et al. 2006; Sembolini et al. 2013; Young et al. 2011; Battaglia et al. 2013; Planelles et al. 2013). By considering a possible time evolution for $\gamma(z)$, such as $\gamma(z)=\gamma_0(1+\gamma_1z)$, and different physical processes in clusters, Planelles et al. (2013) and Battaglia et al. (2013) showed that hot, massive galaxy clusters ($M_{500}>10^{14}$ solar masses) and dynamically relaxed have the following values: $0.55 \leq \gamma_0 \leq 0.79$ and $-0.04 \leq \gamma_1 \leq 0.07$, depending on the physical processes that are included in simulations (see Table 3 in Planelles et al. 2013). For $\gamma_0$ and $\gamma_1$ in non–radiative hydrodynamic simulations, allowed ranges of variation can be taken to be $0.75 \leq \gamma_0 \leq 0.87$ and $-0.04 \leq \gamma_1 \leq 0.07$, respectively. Then, these authors did not find significant dependence of the  $\gamma(z)$ on galaxy cluster redshift. In their measurements these authors  considered $f_{gas}$ as a cumulative quantity into $r_{2500}$. However, the models describing the physics of intra-cluster medium used in hydrodynamic simulations may not span the entire range of physical process allowed by our current understanding. An interesting discrepancy between the observations and the results of the simulations occurs, for instance, with the fraction of stars.  By comparing their results from radiative simulations for stellar fraction in massive galaxy clusters with observations, Planelles et al. (2013) found a larger stellar fraction in massive galaxy clusters, independent of the observational data used in comparison (details can be found in Fig.2 of their paper).}

More recently, Holanda et al. (2017), by assuming the cosmic distance duality relation validity, $D_LD_A^{-1}(1+z)^{-2}=1$ (Etherington 1933; Ellis 2007),  combined  X-ray gas mass fraction measurements in galaxy clusters with luminosity distances, $D_L$, of type Ia supernovae, {  priors on $\Omega_b$ and $\Omega_M$ from Planck results} and reconstructed, via Gaussian processes, the baryon depletion factor up to redshift one\footnote{An interesting test for the cosmic distance duality relation validity by using exclusively gas mass fraction data can be found in Holanda, Gon\c{c}alves \& Alcaniz 2012.}. {  No specific cosmological model was used to obtain distances to galaxy clusters in their methodology}. As result, no observational evidence for a time evolution of the baryon depletion factor was found. The galaxy cluster sample used by these authors was that compiled by Mantz et al. (2014), where the gas mass fractions of galaxy clusters were taken on a  $(0.8-1.2)$ $\times r_{2500}$ shell rather than  integrated at all radii $r \leq r_{2500}$. In this case, the hydrodynamical simulations of Planelles et al. (2013)  showed that the $\gamma_0$ value  presents  only a slight dependence on physical processes and $\gamma_0$ was constrained to be $\gamma_0= 0.85 \pm 0.08$ (see Fig.6 in Planelles et al. 2013). It is important to comment that such approach, where a global argument (the cosmic distance duality relation validity) is used to obtain local properties of astronomical structures, also was considered very recently in strong gravitational lensing observations (Holanda, Pereira \& Deepak 2017) and to infer the elongation of the gas distribution in galaxy clusters (Holanda \& Alcaniz 2017).


\begin{figure*}
\centering
\includegraphics[width=0.32\textwidth]{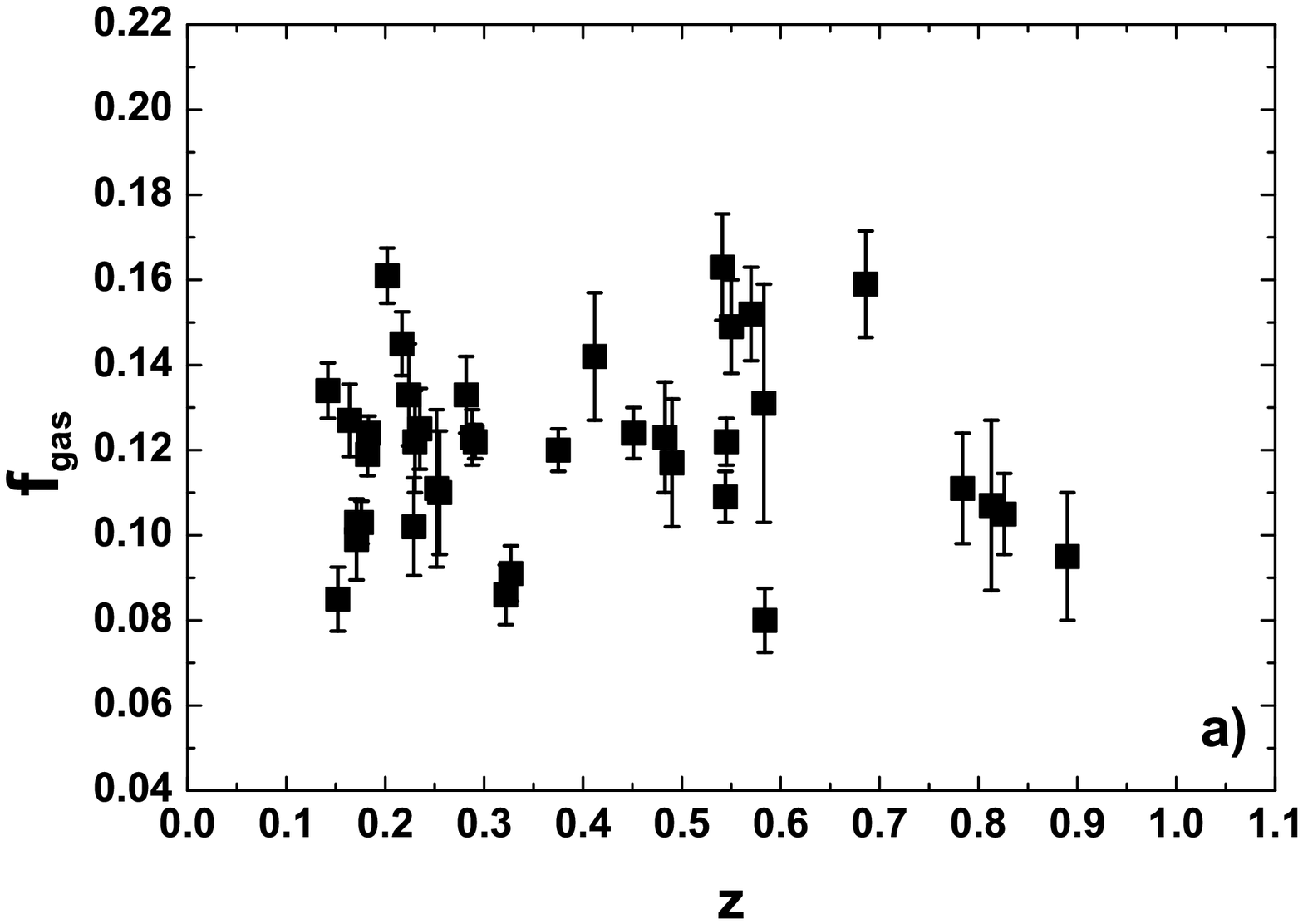}
\includegraphics[width=0.32\textwidth]{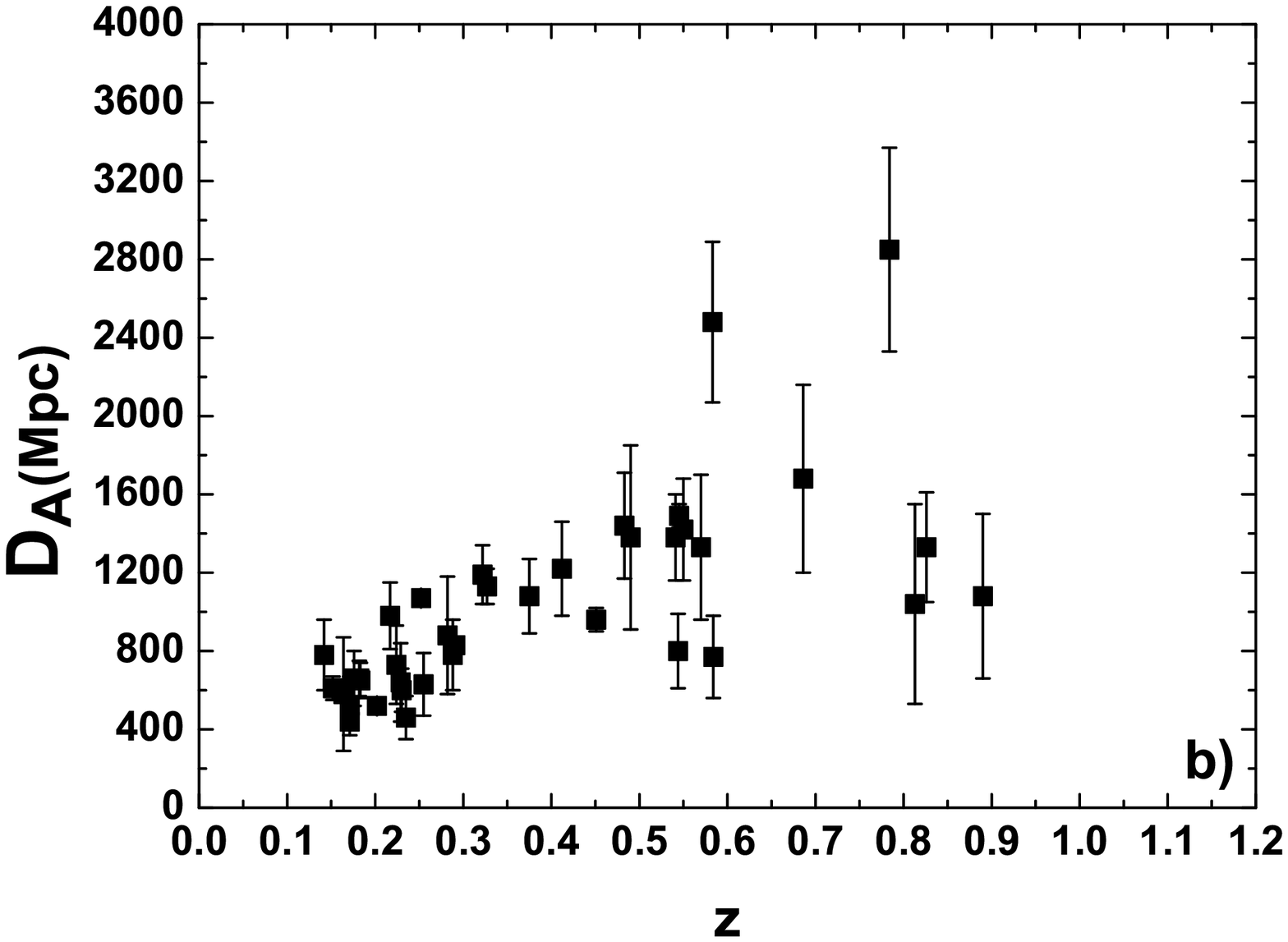}
\includegraphics[width=0.32\textwidth]{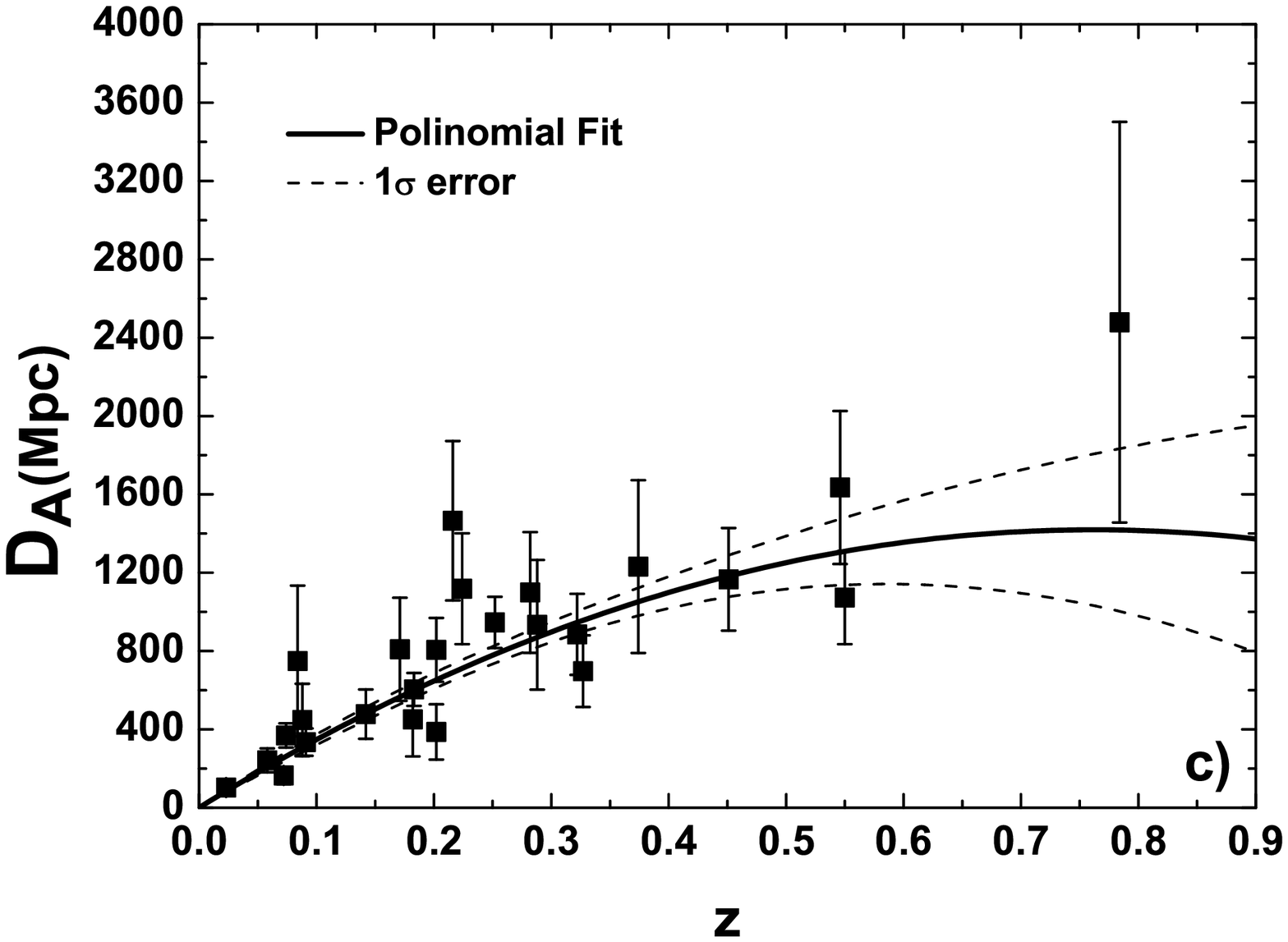}
\caption{Fig.(a) shows the 38 gas mass fractions of galaxy clusters as compiled by La Roque et al.(2006). Fig.(b) shows the 38 angular diameter distances of the same galaxy clusters of  fig.(a) as compiled by Bonamente et al. (2006). Fig.(c) shows the 25 angular diameter distances of galaxy clusters from the De Filippis et al. (2005) sample. }
\end{figure*}

In this paper, we  put  observational constraints on the baryon depletion factor in  galaxy clusters  and its possible time evolution  by using exclusively  galaxy cluster data. The analyses are performed by using two kind of data set from the same galaxy clusters: $f_{gas}$ and  angular diameter distance, $D_A$. The $f_{gas}$ sample corresponds to  38 data from La Roque et al. (2006). The angular diameter distances to these galaxy clusters were obtained through their X-ray surface brightness and Sunyaev-Zel'dovich effect observations by Bonamente et al. (2006). In both data set the galaxy clusters were described by the same electron density and temperature profiles: the non-isothermal spherical double $\beta$-model. In this way, our method to obtain the depletion factor uses the same galaxy clusters, described by the same density and temperature profiles and two kind of observations, where one of them, the $f_{gas}$ measurement, depends on the $\gamma(z)$ value and the other does not. A possible time evolution of the gas mass fraction is  explored by  taking $\gamma(z) = \gamma_0(1+\gamma_1 z)$, where $\gamma_0$ and $\gamma_1$ parametrize the normalization and the time evolution of this quantity, respectively.  We obtain an excellent agreement between our $\gamma_0$ observational value and the simulations, however, a non negligible time evolution is verified, which does not occur in simulations. However, when we consider angular diameter distances from the flat $\Lambda$CDM model (Ade et al. 2015), we obtain $\gamma_0$ and $\gamma_1$ values  in full agreement with cosmological hydrodynamical simulations. {  These conclusions  are also found when we obtain the angular diameter distances  for the galaxy clusters in the La Roque et al. (2006) sample from an angular diameter distance sample of galaxy clusters described by an elliptical profile (De Filippis et al. 2005). Then, our results show that  the non-isothermal double $\beta$-model does not affect the quality of the gas mass fraction measurements of the galaxy clusters considered, but it affects their angular diameter distances.}

The paper is organized as follows. In Section II, we describe the observational quantities. In section III, we discuss the  non-isothermal double $\beta$-model considered in analyses. In Section IV we
 describe the method to obtain $\gamma(z)$. The two galaxy clusters data set are described in Section V. In section
VI, we perform the analyses and present the results. In section VII we discuss our results. Finally,
the conclusions are given in section VIII.

\begin{figure*}
\centering
\includegraphics[width=0.4\textwidth]{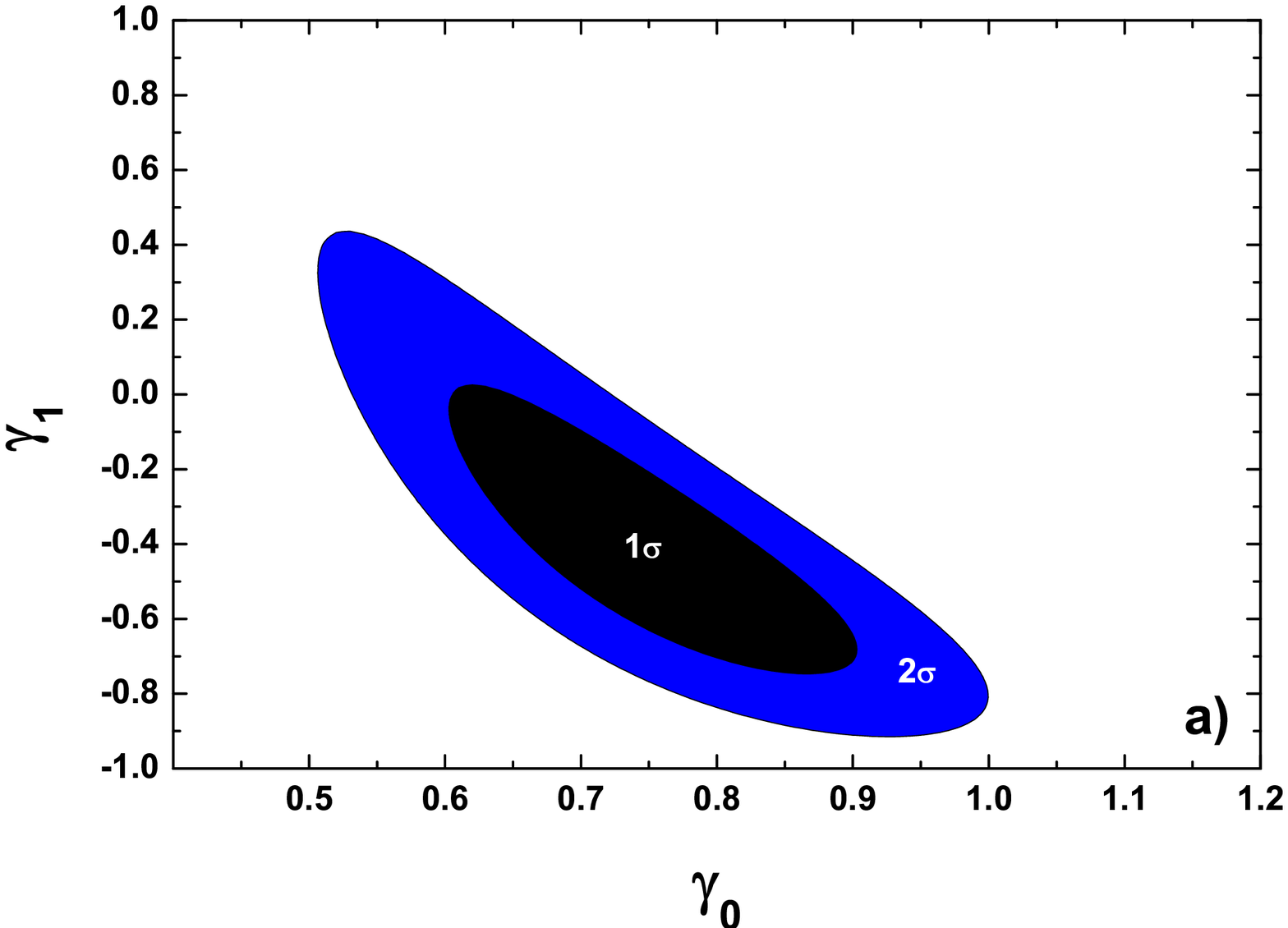}
\includegraphics[width=0.4\textwidth]{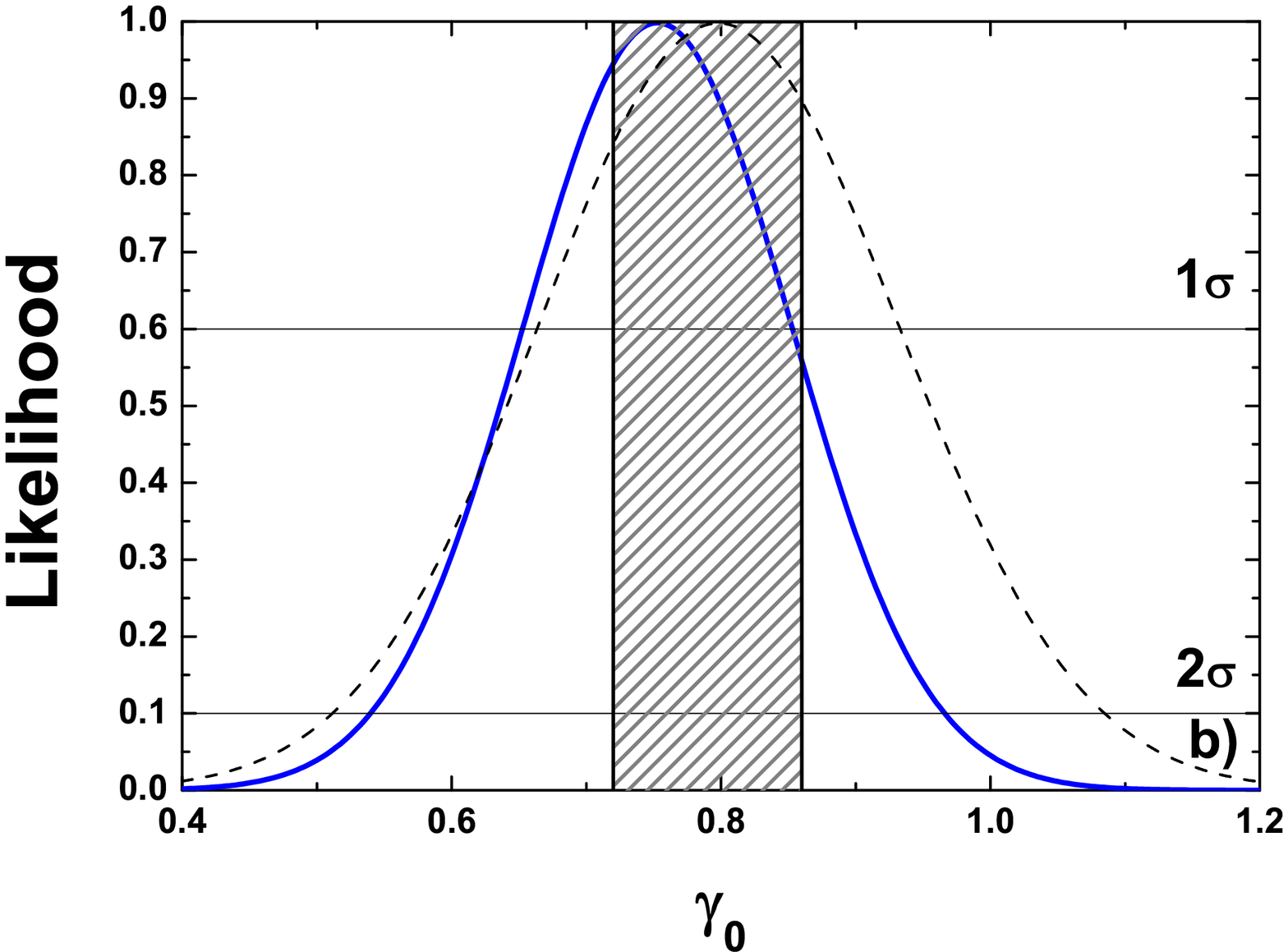}
\includegraphics[width=0.4\textwidth]{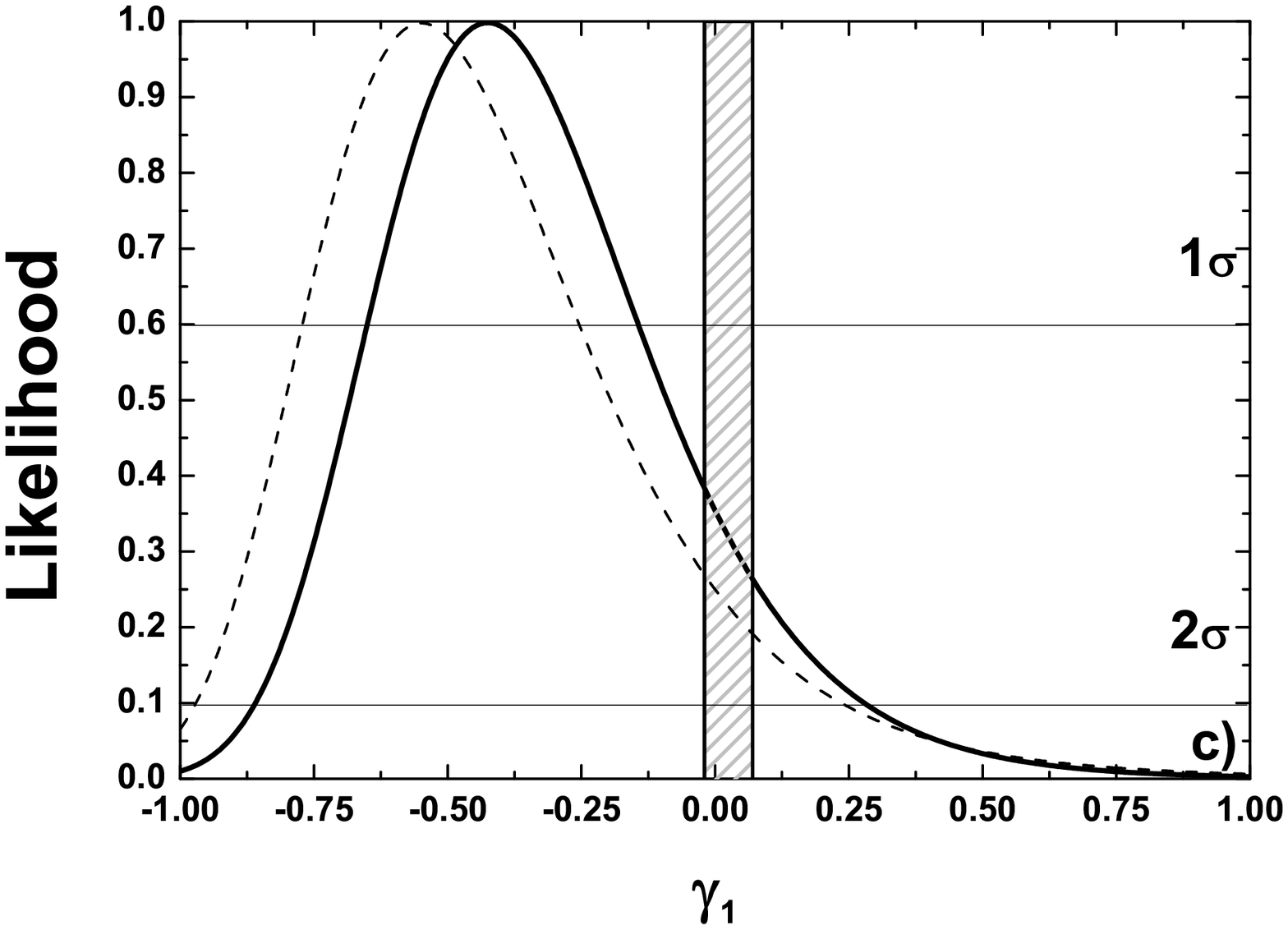}
\includegraphics[width=0.4\textwidth]{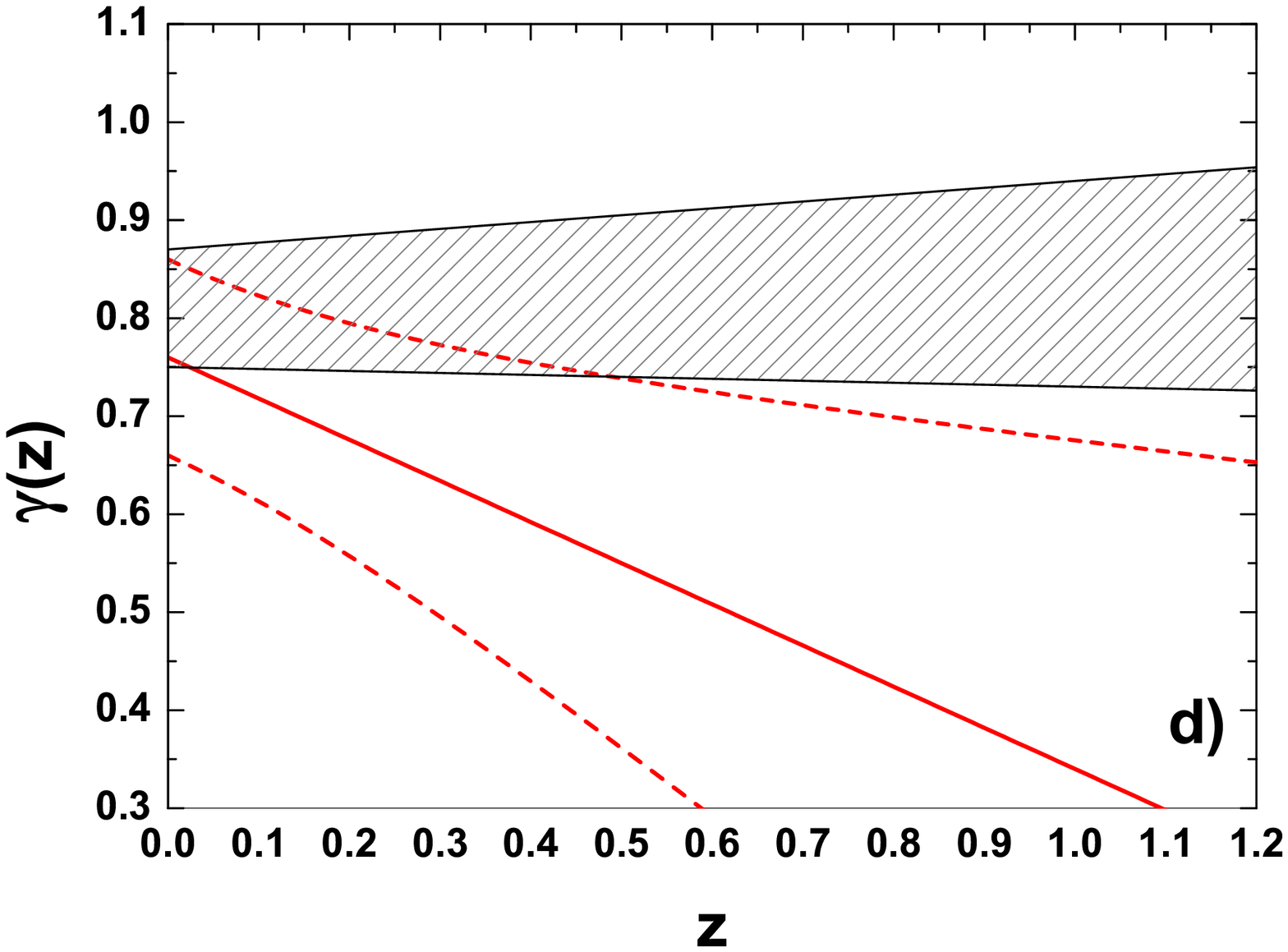}
\caption{Constraints on possible $\gamma(z)$ evolution, such as: $\gamma(z)=\gamma_0(1 + \gamma_1z)$ by using $f_{gas}$ measurements and $D_A$ from Bonamente et al. (2006). Fig.(a) shows the plane for $(\gamma_0, \gamma_1)$. Figs. (b) and (c) show the likelihoods for $\gamma_0$ and $\gamma_1$, respectively, by using the complete samples (solid line) and 29 data points of each sample (dashed line). Fig.(d) shows the evolution for $\gamma_(z)$ by using the 1$\sigma$ results from figs.(b) and (c). The dashed are correspond to results from hydrodynamical simulations.}
\end{figure*}

\begin{figure*}
\centering
\includegraphics[width=0.4\textwidth]{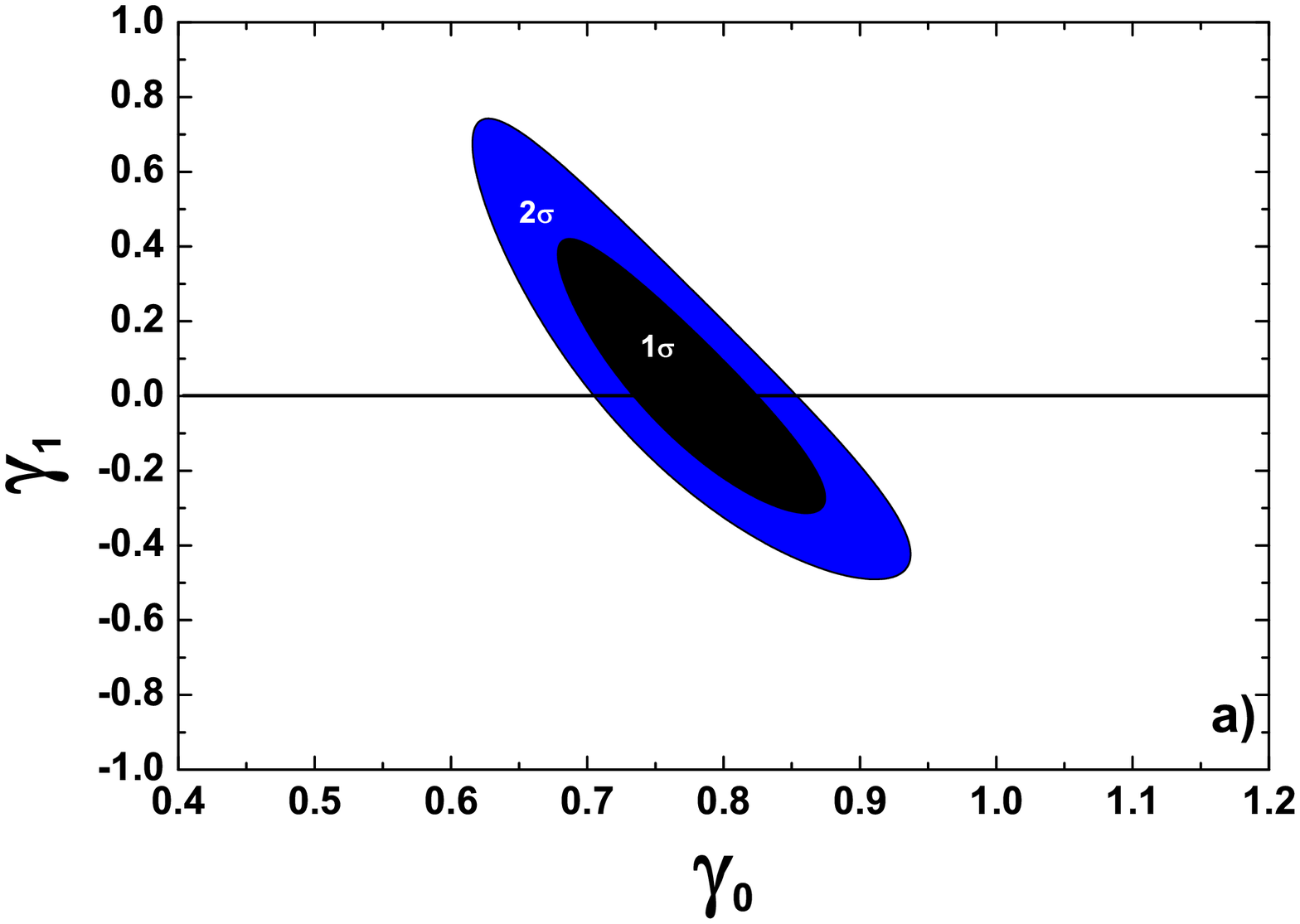}
\includegraphics[width=0.4\textwidth]{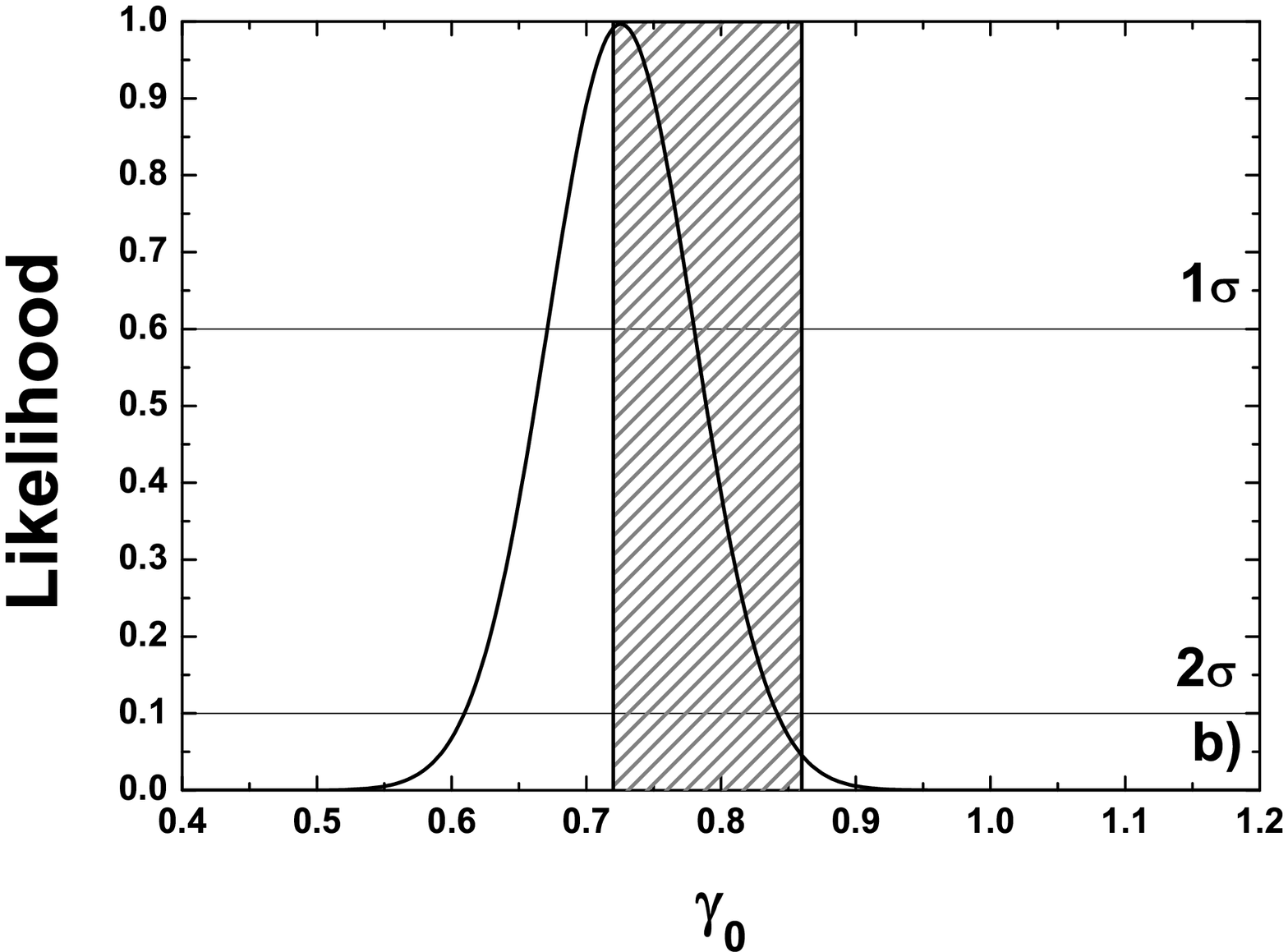}
\includegraphics[width=0.4\textwidth]{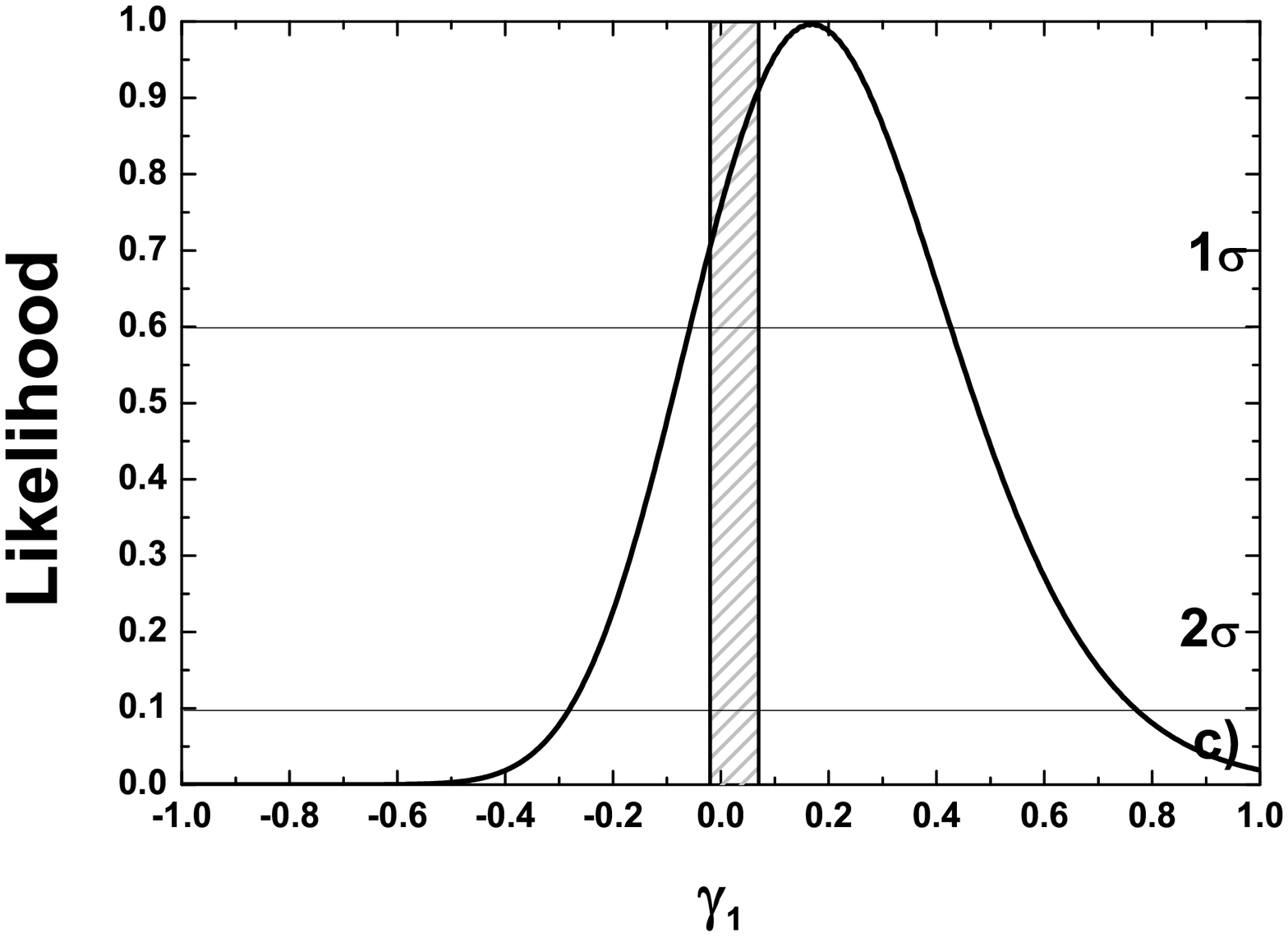}
\includegraphics[width=0.4\textwidth]{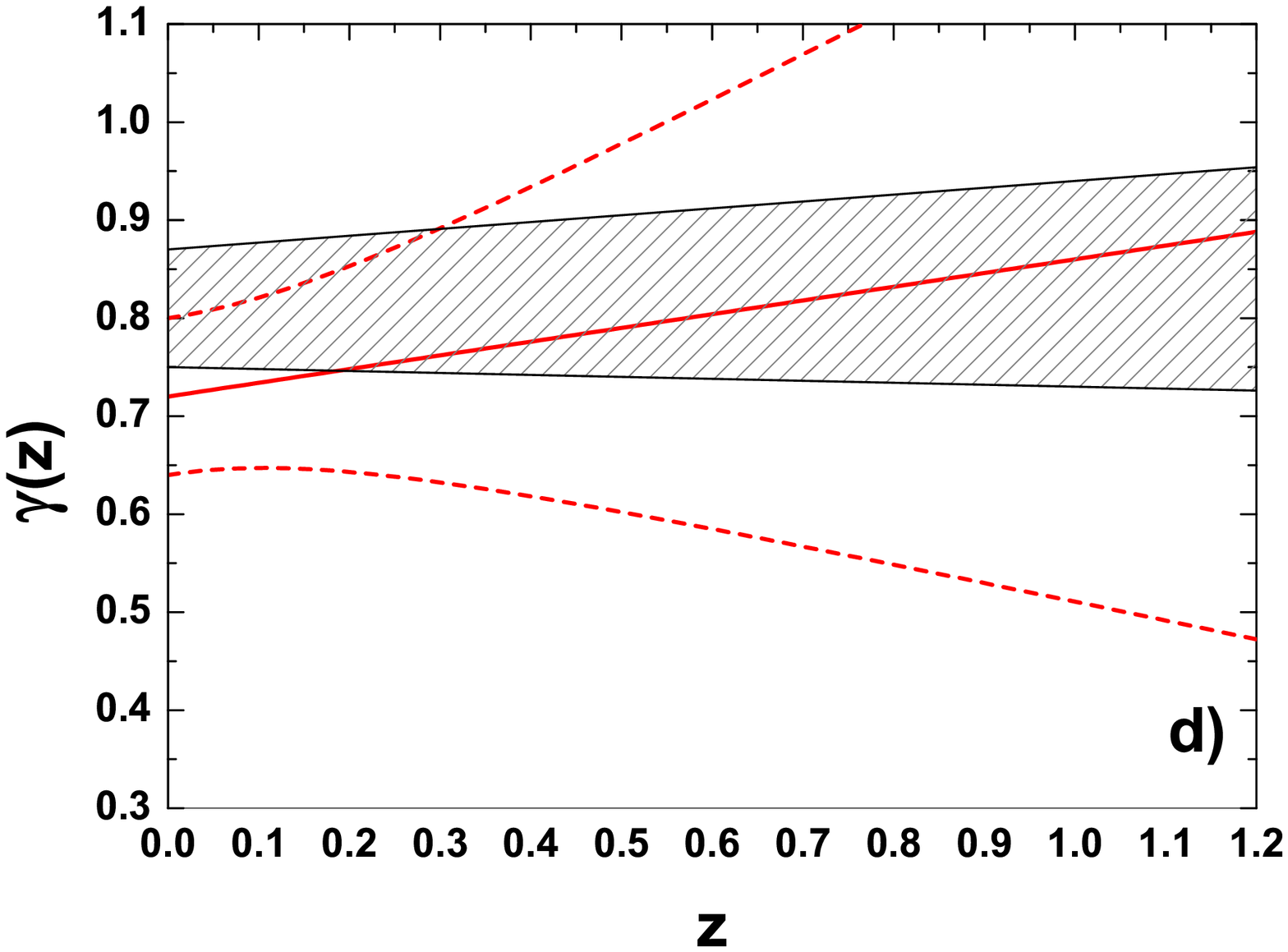}
\caption{Constraints on possible $\gamma(z)$ evolution, such as: $\gamma(z)=\gamma_0(1 + \gamma_1z)$ by using $f_{gas}$ measurements and $D_A$ from  De Filippis et al. (2005). Fig.(a) shows the plane for $(\gamma_0, \gamma_1)$. Figs. (b) and (c) show the likelihoods for $\gamma_0$ and $\gamma_1$, respectively, by using the complete samples (solid line) and 29 data points of each sample (dashed line). Fig.(d) shows the evolution for $\gamma_(z)$ by using the 1$\sigma$ results from figs.(b) and (c). The dashed are correspond to results from hydrodynamical simulations.}
\end{figure*}

\begin{figure*}
\centering
\includegraphics[width=0.4\textwidth]{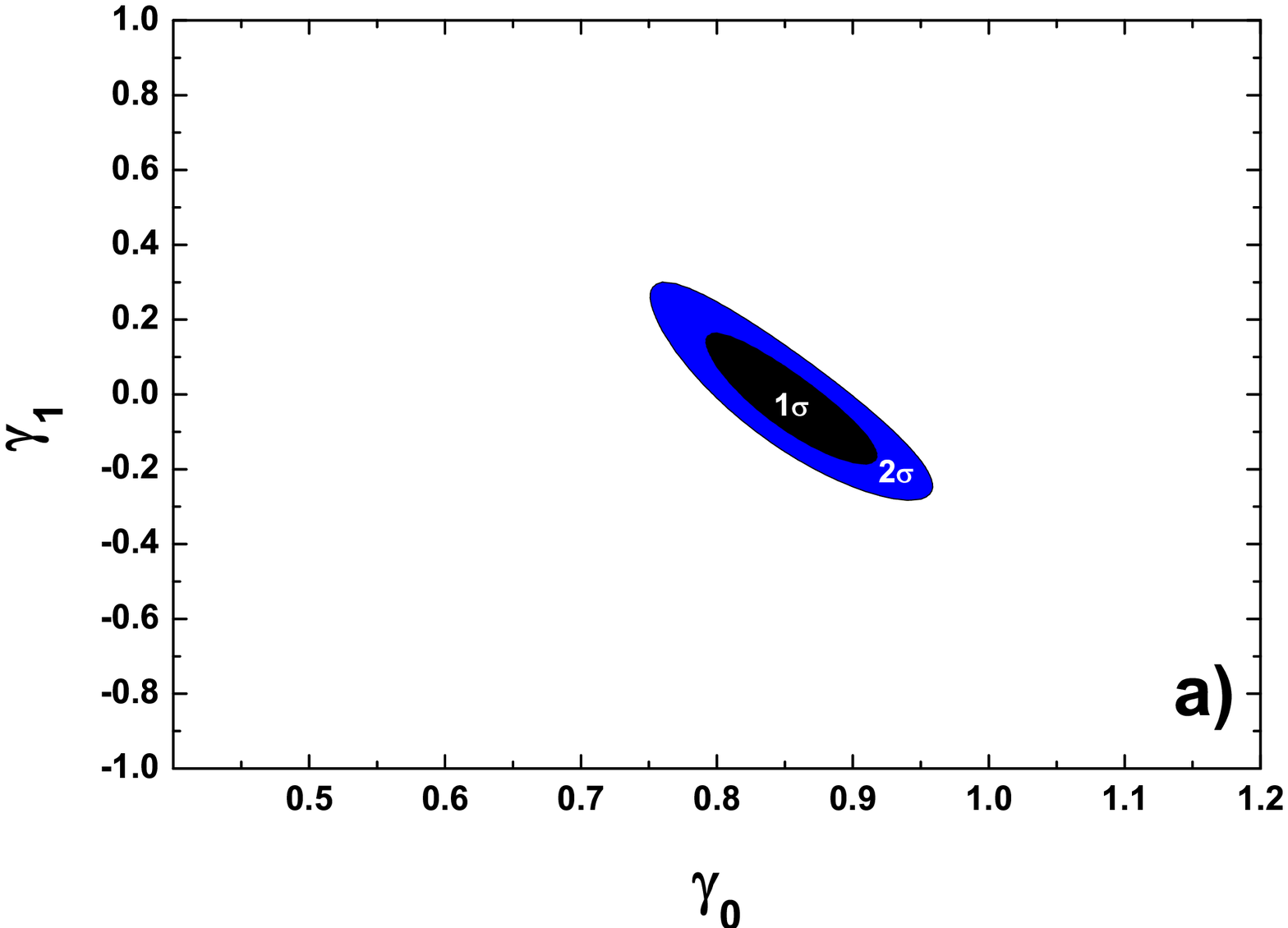}
\includegraphics[width=0.4\textwidth]{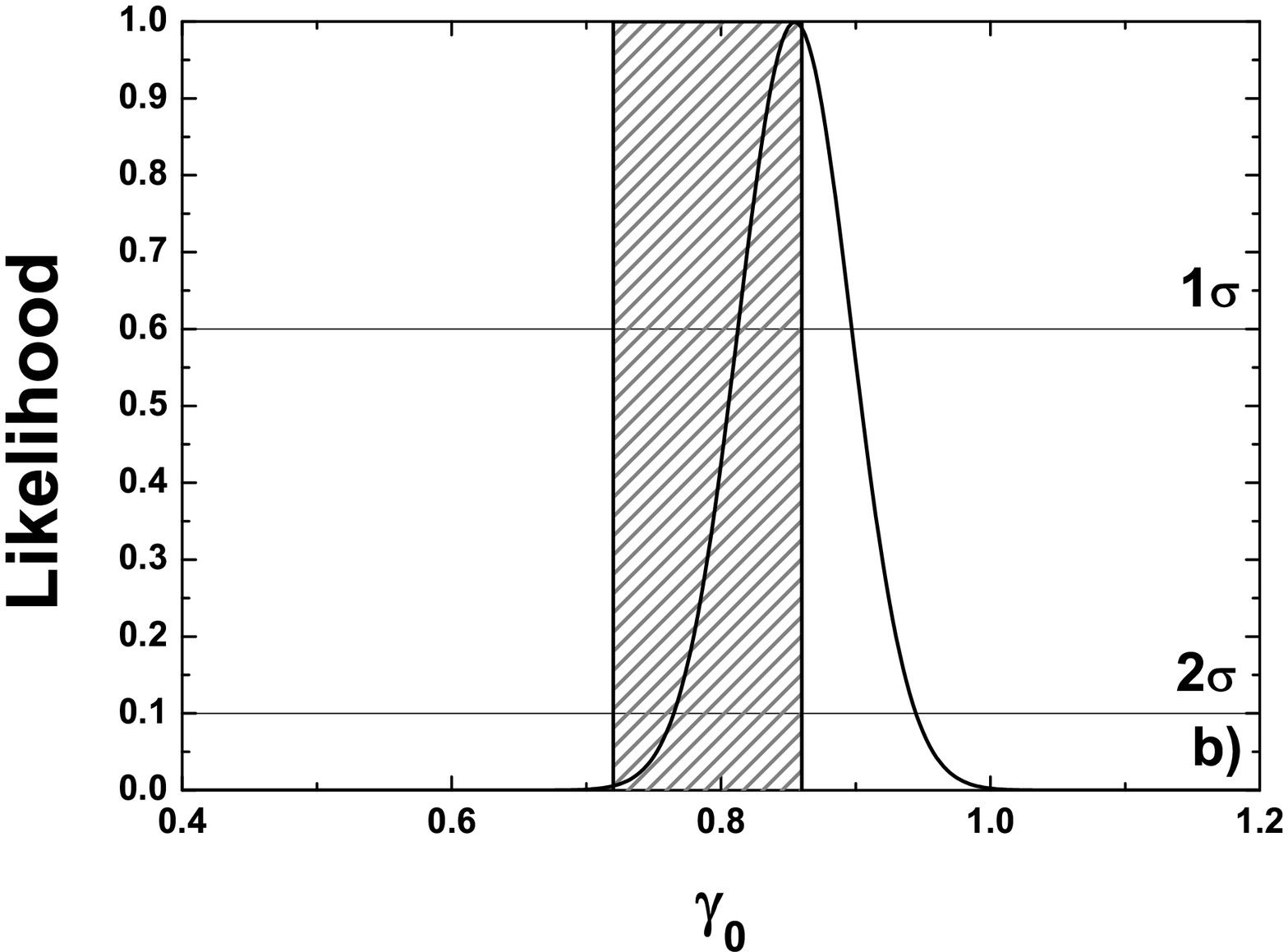}
\includegraphics[width=0.4\textwidth]{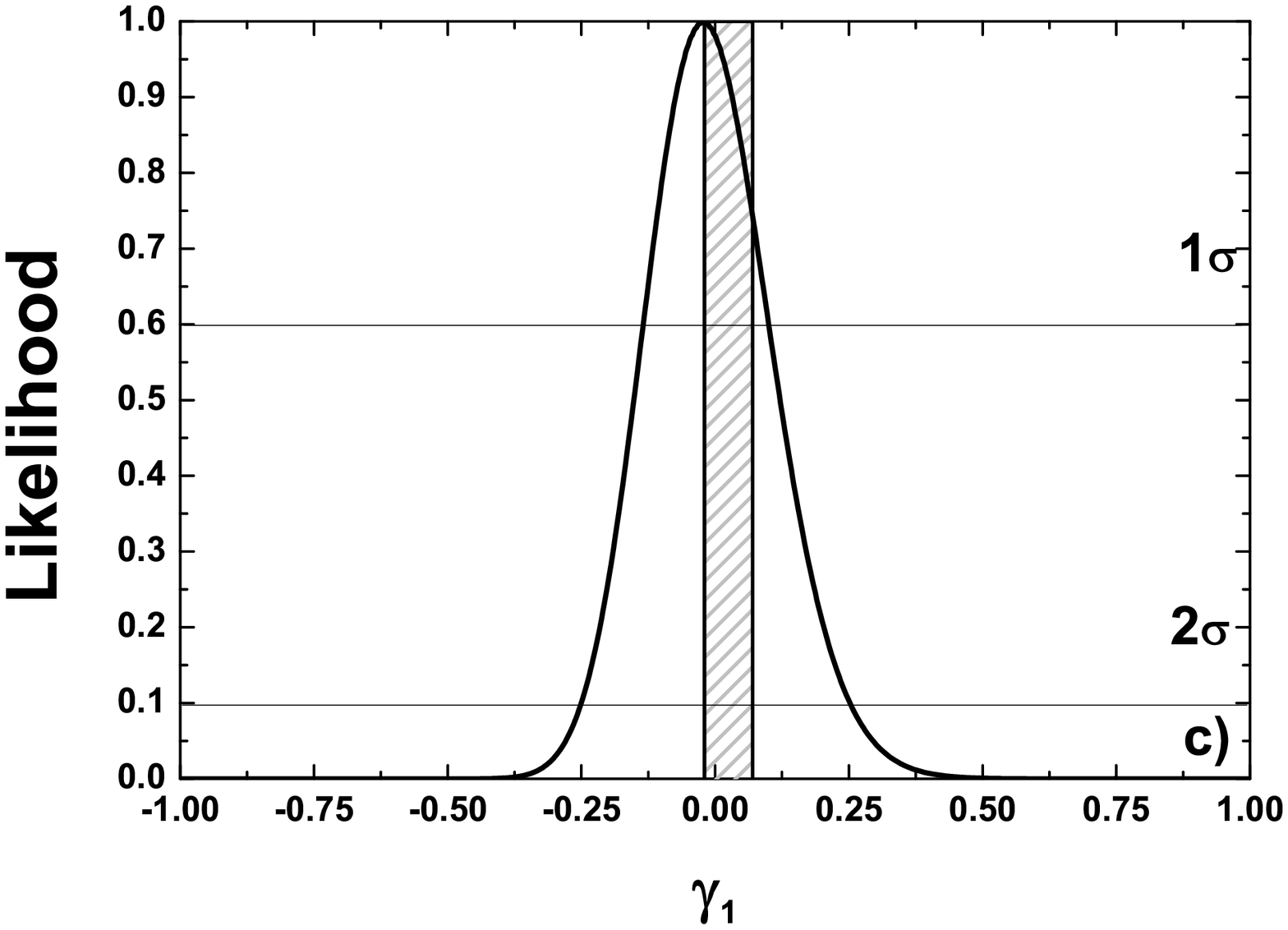}
\includegraphics[width=0.4\textwidth]{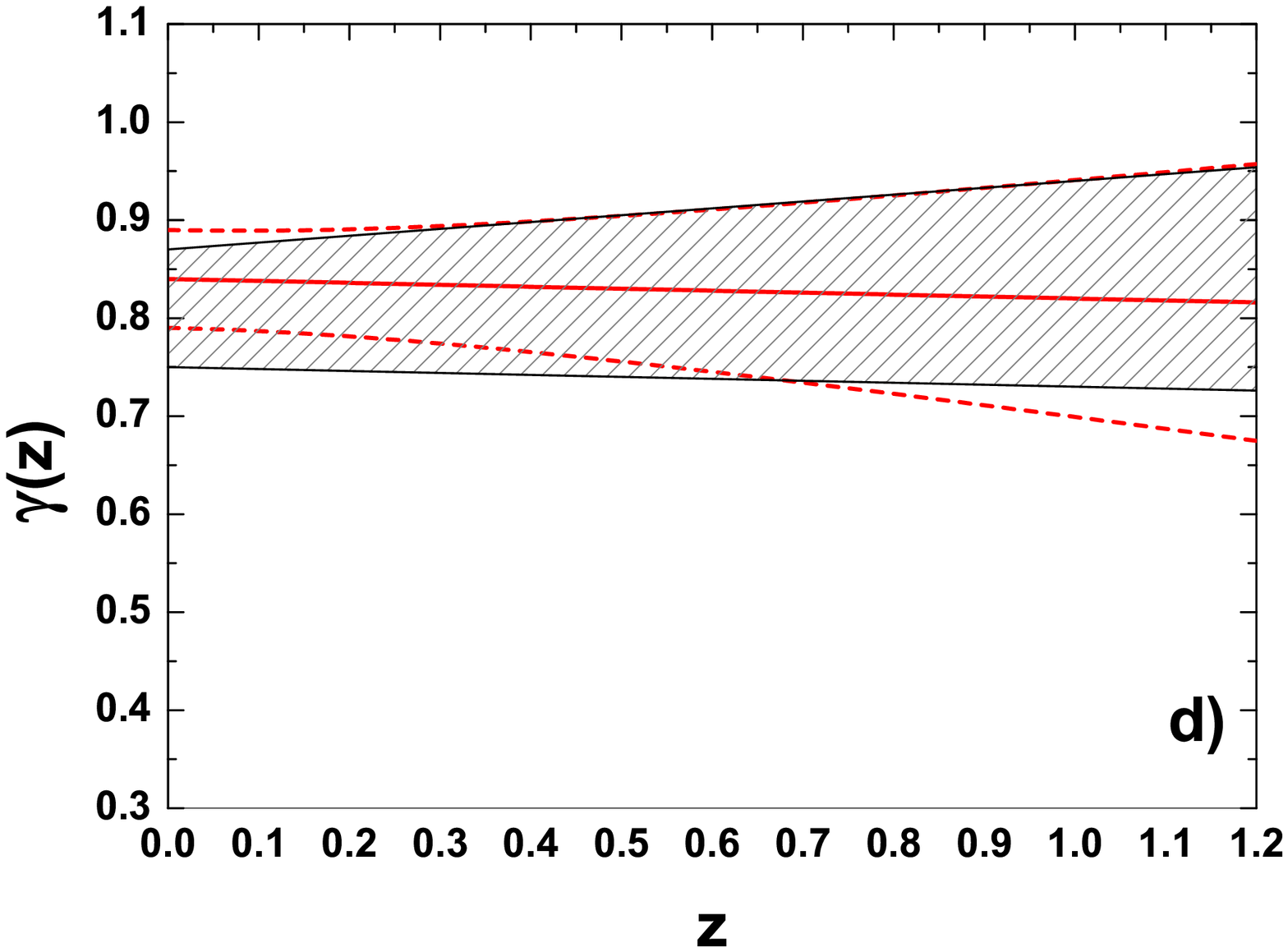}
\caption{Constraints on possible $\gamma(z)$ evolution, such as: $\gamma(z)=\gamma_0(1 + \gamma_1z)$ by using $f_{gas}$ measurements and $D_A$ from Planck results. Fig.(a) shows the plane for $(\gamma_0, \gamma_1)$. Figs.(b) and (c) show the likelihoods for $\gamma_0$ and $\gamma_1$, respectively, by using the complete samples (solid line) and 29 data points of each sample (dashed line). Fig.(d) shows the evolution for $\gamma_(z)$ by using the 1$\sigma$ results from figs.(b) and (c). The dashed are correspond to results from hydrodynamical simulations.}
\end{figure*}
\section{Angular diameter distance and gas mass fraction measurements}

\subsection{Angular diameter distance via SZE and X-ray}

As it is known, since some information about the electron density and temperature profiles of the hot intra-cluster gas of the galaxy clusters be known, it is possible to obtain their angular diameter distances by combining the X-ray surface brightness measurements with the  temperature decrements in the cosmic microwave background spectrum due to thermal Sunyaev-Zel'dovich effect (SZE) (Sunyaev \& Zel'dovich 1972; Carlstrom, Holder \& Reese 2002). Let us  to discuss briefly the SZE/X-ray technique of determining  angular diameter distances of galaxy clusters. 

The equation to SZE is (La Roque et al. 2006)
\begin{equation}
\label{eq:sze1} \frac{\Delta T_{\rm 0}}{T_{\rm CMB}} = f(\nu,
T_{\rm e}) \frac{ \sigma_{\rm T} k_{\rm B} }{m_{\rm e} c^2} \int
_{\rm l.o.s.}n_e T_{\rm e} dl, \
\end{equation}
where  $T_{\rm e}$ is the electronic temperature, $\Delta T_{\rm 0}$ is the central thermodynamic SZE
temperature decrement, $n_e$ is the electronic density of the intra-cluster medium, 
$k_B$ the Boltzmann constant, $T_0$ = 2.728K is the temperature of the cosmic microwave background (CMB), $m_{\rm e}$ the electron mass and $f(\nu,T_{\rm e})$ accounts for frequency shift and relativistic corrections (Itoh, Kohyama \& Nozawa 1998; Nozawa, Itoh \& Kohyama 1998). 

On the other hand, the X-ray emission in galaxy clusters is due to thermal bremsstrahlung and, by considering the intra-cluster medium constituted  by  hydrogen and helium, the X-ray surface brightness can be given by 
\begin{equation}
S_x = \frac{1}{4\pi (1+z)^4} \! \int \!\!  \, n^2_e  \LameH \,d\ell
\label{eq:xray_sb}
\end{equation}
where the integral is along the line of sight and $\LameH$ is the X-ray cooling function, proportional to $T_e^{1/2}$ (gas temperature) (Sarazin 1988).  

Then, by assuming the classical isothermal $\beta$-model for the electron density of the hot
intra-cluster gas (Cavaliere \& Fusco-Fermiano 1978)
\begin{equation}\label{eqprof}
 n_e(r) =
 n_0\left[1+\left(\frac{r}{r_c}\right)^2\right]^{-3\beta/2},
\end{equation}
 one can solve equations (\ref{eq:sze1}) and (\ref{eq:xray_sb}) for the angular diameter
distance: 

\begin{eqnarray}
\label{eqobl7}
{{D}}_A(z) &= & \left[ \frac{\Delta {T_0}^2}{S_{\rm X0}}
\left( \frac{m_{\rm e} c^2}{k_{\rm B} T_{e} } \right)^2
\frac{g\left(\beta\right)}{g(\beta/2)^2\ \theta_{\rm c}}
\right] \times \nonumber \\
& & \times \left[ \frac{\Lambda_e }{4 \pi^{3/2}f(\nu,T_{\rm
e})^2\ {(T_0)}^2 {\sigma_{\rm T}}^2\ (1+z)^4} \right] \nonumber \\,
\end{eqnarray}
with
\begin{equation}
g(\alpha)\equiv\frac{\Gamma \left[3\alpha-1/2\right]}{\Gamma \left[3
\alpha\right]}, \label{galfa}
\end{equation}
where $S_{x0}$ is the central surface brightness, $\beta$ determines the slope at large radii, $r_c$ is the core radii ($\theta_c=r_c/D_A$) and $ \Gamma(\alpha)$ is the gamma function and $z$ is the galaxy cluster redshift. This technique for measuring distances is completely independent of other methods  and it can be used to measure distances at high redshifts directly.

\subsection{X-ray gas mass fraction}

The gas mass fraction in galaxy clusters is defined as $f_{gas} =M_{gas}/M_{Tot}$, where  $M_{gas}$ is the gas mass obtained by  integrating a electron density profile and  $M_{Tot}$ is the total mass obtained via hydrostatic equilibrium assumption. Then, under this assumption and isothermality, $M_{Tot}$ and $M_{gas}$ are given by (Sasaki 1996; Grego et al. 2001; La Roque et al. 2006)
\begin{equation}
M_{Tot}(r)=\frac{3\beta kT_e D_A}{G \mu m_p} \frac{\theta^3}{\theta_c^2 + \theta^2},
\label{eq:mtot_hse_iso}
\end{equation}
 and
\begin{equation}
M_{gas}(r) = \varsigma \int_{0}^{r/\Da} \left (1+\frac{\theta^2}{\theta_c^2}
\right)^{-3\beta/2}\: \theta^2 d\theta,
\label{eq:mgas_single}
\end{equation}
where here was considered the profile (\ref{eqprof}),  $\varsigma=4\pi \mu_e n_{e0} m_p\, D_A^3$, $\mu_e$ and $m_p$ are the mean molecular weight of the electrons and the proton mass, respectively. Then, if a given cosmological model is assumed for angular diameter distance ($D_A$),  the model central electron density, $n_{e0}$, can be  analytically obtained from equations (2) and (3)  (La Roque et al. 2006)
\begin{equation}
n^{X-ray}_{e0} = \left( \frac{\Xo \:4\pi (1+z)^4\:
  \Gamma(3\beta)}{\LameH D_A \pi^{1/2}\: \Gamma(3\beta-\frac{1}{2})\,
  \theta_c} \right)^{1/2}.
\label{eq:xray_ne0}
\end{equation}
So, it is possible to obtain the well-known relation $f_{X-ray} \propto {D_A}^{3/2}$. As one may see, the $f_{gas}$ measurements are dependent on the cosmological model adopted in observations.

\section{The non-isothermal double $\beta$-model}

Through decades, the isothermal $\beta$-model was used in the analysis of X-ray and Sunyaev-Zel'dovich effect galaxy cluster images (Cavaliere \& Fusco-Femiano 1976, 1978; Jones \& Forman 1984; Elbaz et al. 1995; Grego et al. 2001; Reese et al. 2002; Ettori et al. 2004). However, this profile  fails to provide a good description of the X-ray surface brightness observed in some galaxy cluster cores, more precisely, in highly relaxed clusters with sharply peaked central X-ray emission: the cool core galaxy clusters. Actually, cool core clusters effectively exhibit two components: a centrally concentrated gas peak and a broad, shallower distribution of the gas. In this way, a double $\beta$-model of the surface brightness was first used by Mohr et al. (1999)  and further developed by La Roque et al. (2006). In this case, the 3-D density profile of the non-isothermal spherical double-$\beta$ model is expressed by,
\begin{equation}
n_e({\mathbf{r}}) = n_{e0} \left[ f\: \!\left ( 1 +
\frac{r^2}{\,r_{c_1}\!\!\!\!^2}\right) ^{-3\beta/2} + (1-f)\:\left (1 +
\frac{r^2}{\,r_{c_2}\!\!\!\!^2}\right)^{-3\beta/2}\right]
\label{eq:double_beta}
\end{equation}
where the narrow, peaked central density component and the broad, shallow outer density profile, respectively, are described by the two core radii, $r_{c_1}$ and $r_{c_2}$. The fractional contribution of the narrow, peaked component to the central density is given by the factor $f$. If $f=0$  the isothermal $\beta$-model is recovered (Eq.3). Here, in order to reduce the total number of degrees of freedom, the same $\beta$ is used for both the central and outer distributions.

By assuming  the hydrostatic equilibrium {  ()} and spherical one obtains 
\begin{equation}
\frac{dP}{dr}=-\rho_g \frac{d\phi}{dr},
\end{equation}
where $P$, $\rho_g$ and $\phi=-G M(r)/r$ are the gas pressure,  gas density and  the gravitational potential due to dark
matter and the plasma. By considering the ideal gas equation of state for
the diffuse intra-cluster plasma, $P=\rho_g k_B T / \mu \; m_p$,
where $\mu$ is the total mean molecular weight and $m_p$ is the proton
mass,  a relationship between the cluster temperature and
its mass distribution can be find,
\begin{eqnarray}
\label{eq-hse}
\frac{dT}{dr} &=&-\left( \frac{\mu m_p}{k_B} \frac{d\phi}{dr} +
\frac{T}{\rho_g} \frac{d\rho_g}{dr} \right), \nonumber \\
&=& -\left( \frac{\mu m_p}{k_B}
\frac{G M}{r^2} + \frac{T}{\rho_g} \frac{d\rho_g}{dr} \right).
\end{eqnarray}
{Moreover, Bonamente et al.\ (2006) and La Roque et al. (2006) combined hydrostatic equilibrium equations
with a dark matter density distribution from Navarro, Frank and
White (1997),
\begin{equation}
\rho_{\scriptscriptstyle DM}(r)=N \left[ \frac{1}{(r/r_s) (1+r/r_s)^2}
\right],
\label{NFW}
\end{equation}
where $N$ is a density normalization constant and $r_s$ is
a scale radius. {  these authors obtained the free parameters of these equations ($n_{e0}$, $f$,
$r_{c1}$, $r_{c2}$, $\beta$, $\mathcal{N}$ and $r_s$) for 38 galaxy clusters by calculating a joint likelihood of the X-ray and
SZE data by using a Markov chain Monte Carlo method proposed by Bonamente et al. (2004). The galaxy clusters are the same in both sample. As one may see, the cluster plasma and dark matter distributions were analyzed by assuming spherical symmetry and hydrostatic equilibrium model, thereby accounting for radial variations in density, temperature and abundance.  The exact values of the angular diameter distances and gas mass fractions used in our analysis can be found on the Table 2 of Bonamente et al. (2006) and Table 5 of La Roque et al. (2006), respectively (see section V for details).}}

\section{Method}

The gas mass fraction in galaxy clusters can be linked to cosmic baryon fraction and used as a cosmological tool via the following equation (Sasaki 1996; Allen et al. 2008; Mantz et al. 2014):

\begin{eqnarray} \label{eq:fgasmodel}
  f_{gas} \left(z\right) \hspace{3.4cm}\nonumber\\
  \hspace{4em}= A(z) K(z) \,  \gamma(z) \left( \frac{\Omega_b}{\Omega_M} \right) \left[ \frac{D_A^\mathrm{ref}(z)}{D_A(z)} \right]^{3/2},
\end{eqnarray}
where $\Omega_b$ is the baryon density parameter, the $K(z)$ parameter quantifies inaccuracies in instrument calibration, as well as any bias in measured masses due to substructure, bulk motions and/or non-thermal pressure in the cluster gas. Recently, Apllegate et al. (2016)  calibrated {\it Chandra} hydrostatic masses to relaxed clusters with accurate weak lensing measurements from the Weighing the Giants project. The $K(z)$ parameter was obtained to be $K=0.96 \pm 0.09$ and no significant trends  with mass, redshift or the morphological indicators was verified. The ratio into brackets computes the expected variation in $ f_{gas}$ when the underlying cosmology is varied. The index "`ref"' corresponds to fiducial cosmological model used to obtain the $f_{gas}$. {  The term $D^{ref}_A$ removes all the dependence of the gas mass fraction  with respect to the reference cosmological model used in the observations.} $\gamma(z)$ is the quantity of our interest: the baryon depletion factor. There is yet a $A(z)$ factor, corresponding to angular correction factor, which is close to unity for all cosmologies and redshifts of interest and it can be neglected without significant loss of accuracy for most work. In this way, from Eq.(\ref{eq:fgasmodel}) one obtains:

\begin{equation} \label{gamma}
\gamma(z)=\frac{f_{gas}}{K(\Omega_b/\Omega_M)}\left(\frac{D_A}{D_A^\mathrm{ref}}\right)^{3/2}.
\end{equation}
In order to perform our analyses, we use {\it Planck} priors  on $\Omega_b$ and $\Omega_M$ (see table I). Unlike Holanda and co-workers (2017), which used the cosmic distance duality relation validity and replaced the $D_A$ quantity by $D_L$ of SNe Ia, the $D_A$ quantity  for each galaxy cluster used in our analyses was obtained from the Sunyaev-Zel'dovich and X-ray observations and compiled by Bonamente et al. (2006). The $D_A^\mathrm{ref}$ from fiducial model is obtained by using the standard equation for the flat $\Lambda$CDM model:

\begin{equation}
\label{ADDLCDM}
D^{ref}_A = \frac{c H_0^{-1}}{(1+z)} \int^{z}_{0}{\frac{dz'}{\sqrt{\Omega_{M} (1+z')^{3} + (1-\Omega_{M})}}} \; {\rm{Mpc}}\;,
\end{equation}
where  $c$ is the speed of light. In the fiducial model: $H_0=70$km/s/Mpc and $\Omega_M=0.3$. 

\begin{table}
\caption{A summary of the priors used in this paper ($h=H_0/100$) (Ade et al. 2015).}
\label{table1}
\par
\begin{center}
\begin{tabular}{|c||c|c|c|}
\hline\hline  & $H_0$ (km/s/Mpc) & $\Omega_b h^2$ & $\Omega_M$   
\\ \hline\hline 
{\it Planck} & $67.74 \pm 0.46$ & $0.02230 \pm 0.00014$ & $0.3089 \pm 0.0062$
\\
\hline\hline
\end{tabular}
\end{center}
\end{table}

\section{Samples}

The samples used in our analyses are:

\begin{itemize}
\item 38 $f_{gas}$ measurements in redshift range $0.14 \leq z \leq 0.89$ from La Roque et al. (2006).{  All galaxy clusters in this sample have total mass higher than $10^{14}$ solar masses and $T_X \geq 5 keV$ (hot galaxy clusters).} These measurements were derived from Chandra X-ray data and OVRO/BIMA interferometric SZE measurements. As commented earlier, the gas density was modeled by a non-isothermal double $\beta$-model, which generalizes the single $\beta$-model profile. A motivation for considering this model is to assess the biases arising from the isothermal assumption and the effects of the cool cores presence. Moreover, the clusters were analyzed by assuming the hydrostatic equilibrium model with the dark matter density distribution profile given by Navarro, Frenk \& White (1996), spherical symmetry with radial variations in density, temperature and abundance being considered. The $f_{gas}$ measurements were obtained within $r_{2500}$ (see Fig.1a). 
\item 38 $D_A$ from the same galaxy clusters of La Roque et al. (2006). These measurements were compiled by Bonamente et al. (2006) by using  Chandra X-ray data and OVRO/BIMA interferometric SZE measurements. The assumptions on the temperature, gas and dark matter profiles are the same adopted by La Roque et al. (2006). It is very important to comment that the SZE/X-ray technique for measuring distances is completely independent of $\gamma$ parameter (see Fig.1b). 
\item {  25 $D_A$ obtained via Sunyaev-Zeldovich effect + X-ray surface brightness observations from a sample compiled by De Filippis et al. (2005) which contains  galaxy clusters in the redshift range $0.023 \leq z \leq 0.784$ (see Fig.1c). De Fillipis et al. (2005) used an isothermal elliptical 2-dimensional $\beta$-model to describe the galaxy clusters. As discussed by these authors, the choice of circular rather than elliptical model affects significantly the values to core radius ($\theta_c$) and, consequently, to the $D_A$, since $D_A\propto \theta^{-1}_c$. {  The exact values of the angular diameter distances used in our analysis can be found on the Table 3 of De Filippis et al. (2006).

However, the galaxy clusters in the  La Roque et al. (2006) and De Filippis et al. (2005) samples not are in identical redshifts. As it is necessary galaxy clusters on the same redshift, we fit the angular diameter distances from De Fillipis et al. (2005) data with a second degree polynomial fit, $D_A(z)= Az + Bz^2 $, where (in Mpc), $A=3725.33107  \pm 350.22694$, $B=-2444.71723 \pm 1026.93663$ and $cov(A,B)=-312713.0836$, with $\chi^2_{red} = 1.1$. In this case, when a third degree polynomial fit is performed, the additional term is compatible with zero. The 1$\sigma$ error from polynomial fits is given by: 

\begin{eqnarray}
\sigma^2 & = & \left(\frac{\partial D^c_A}{\partial A}\right)^2\sigma^2_A +\left(\frac{\partial D^c_A}{\partial B}\right)^2\sigma^2_B \\ & & \nonumber + 2\left(\frac{\partial D^c_A}{\partial A}\frac{\partial D^c_A}{\partial B}\right)cov(A,B).                                                        
\end{eqnarray}}}
\end{itemize}

\section{Analyses and results} 

Our statistical analyses is performed by defining the likelihood distribution function ${\cal{L}} \propto e^{-\chi^{2}/2}$, with
\begin{equation}
\label{chi2} 
\chi^{2} = \sum_{i = 1}^{38}\frac{{(\gamma(z)- \gamma_{obs}(z) })^{2}}{\sigma^{2}_{i, obs}},
\end{equation}
where $\gamma(z)=\gamma_0(1 + \gamma_1z)$ {  is the quantity of our interest}, with $\gamma_0$ being the normalization factor and $\gamma_1$  parametrize a possible time evolution with $z$. {  The quantity $\gamma_{obs}(z)$ is given by Eq.(14) by using the samples of section V, the priors of Table I and the Eq.(15).}The $\sigma^{2}_{i, obs}$ stands for the statistical errors of $K$, $\Omega_b$, $\Omega_M$, $D_A$ and $f_{gas}$.  The common statistical contributions to $f_{gas}$ are: SZE point sources $\pm 4\%$,  kinetic SZ $\pm 8\%$, $\pm 20\%$  and $\pm 10\%$ for cluster asphericity to $f_{gas}^\mathrm{X-ray}$. The common statistical error contributions  for $D_A$ are: SZE point sources $\pm8\%$, X-ray
background $\pm2\%$, Galactic NH $\pm  1\%$, $\pm 15\%$ for cluster asphericity, $\pm8\%$ kinetic SZ and for CMBR anisotropy $\pm2\%$.

The asymmetric error bars present in La Roque et al. (2006) and Bonamente et al. (2006) papers were treated as discussed in D'Agostini (2004) , i. e., $f_{gas} =  \widetilde{f}_{gas} + \Delta_+ - \Delta_-$, with $\sigma_{f_{gas}} = (\Delta_+ +\Delta_-)/2$, where  $\widetilde{f}_{gas}$ stands for the La Roque et al. (2006) measurements and $\Delta_+$ and $\Delta_-$ are, respectively, the associated upper and lower errors bars. The same treatment was performed on the $D_A$ data.

\subsection{Results by using $f_{gas}$ and $D_A$ from Bonamente et al. (2006)}

{  Our statistical results are plotted in fig.(2). Fig.(2a) shows the regions for 1$\sigma$ ($\Delta \chi^2=2.30$) and 2$\sigma$ ($\Delta \chi^2=6.17$) on the $(\gamma_0,\gamma_1)$ plane for two free parameters. For this case, we obtain at 1$\sigma$: $\gamma_0=0.76 \pm 0.14$ and  $\gamma_1=-0.42^{+0.42}_{-0.40}$. Fig.(2b) shows the likelihood for $\gamma_0$ by marginalizing on $\gamma_1$ and Fig.(2c) shows the likelihood for $\gamma_1$ by marginalizing on $\gamma_0$. We obtain at 1$\sigma$ ($\Delta \chi^2=1$): $\gamma_0 = 0.76 \pm 0.10$ and  $\gamma_1 = -0.42^{+0.32}_{-0.30}$. In these two last panels, the 2$\sigma$ interval corresponds to ($\Delta \chi^2=4.0$).}The dashed areas show the results from simulations: $\gamma_0=0.81 \pm 0.06$ and $−0.02 \leq \gamma_1 \leq 0.07$ (Planelles et al. 2013). As one may see, the $\gamma_0$ value is in full agreement with cosmological hydrodynamical simulations. On the other hand, our $\gamma_1$ value is significantly negative, which indicates a time evolution of the gas fraction. Finally,  Fig.(2d) shows the $\gamma(z)$ evolution (obtained by using the  1$\sigma$ results  from Figs.2b and 2c)  and the results from hydrodynamical simulations (dashed area). As one may see, the $\gamma(z)$ function shows  a not negligible time evolution for gas mass fraction in galaxy clusters. 

However,  some galaxy clusters present questionable  reduced $\chi^{2}$ ($2.43 \leq \chi_{d.o.f.}^2 \leq 41.62$) when described by the non-isothermal double $\beta$ model  (see Table 6 in Ref. Bonamente et al. 2006). They are: Abell 665, ZW 3146, RX J1347.5-1145, MS 1358.4 + 6245, Abell 1835, MACS J1423+2404, Abell 1914, Abell 2163, Abell 2204. We excluded these clusters and the results are plotted in Figs.(2b) and (2c) as dashed lines. As one may see, even after excluding these data, the results still show a non-negligible time evolution for $\gamma(z)$. 

These results are in conflict with those from Holanda et al. (2017), where a constant $\gamma(z)$  was obtained by using 40 X-ray emitting gas mass fraction measurements and luminosity distance measurements from SNe Ia. In their analyses, the galaxy clusters were  described under the assumptions of spherical symmetry and hydrostatic equilibrium, but without considering the non-isothermal double $\beta$-model. Moreover, the $f_{gas}$ measurements were taken on a  $(0.8-1.2)$ $\times r_{2500}$ shell rather than  integrated at all radii $r \leq r_{2500}$. In this way, these results show either a strong tension between the non-isothermal double $\beta$-model and a constant $\gamma(z)$ function or really they indicate a time evolution for $\gamma(z)$.

\subsection{Results by using $f_{gas}$ and $D_A$ from De Filippis et al. (2005)}

For this case, our statistical results are plotted in fig.(3). Fig.(3a) shows the regions for 1$\sigma$ ($\Delta \chi^2=2.30$) and 2$\sigma$ ($\Delta \chi^2=6.17$)  on the $(\gamma_0,\gamma_1)$ plane for two free parameters. For this case, we obtain at 1$\sigma$: $\gamma_0=0.72 \pm 0.01$ and  $\gamma_1=0.16 \pm 0.36$. Fig.(2b) shows the likelihood for $\gamma_0$ by marginalizing on $\gamma_1$ and Fig.(2c) shows the likelihood for $\gamma_1$ by marginalizing on $\gamma_0$. We obtain at 1$\sigma$  ($\Delta \chi^2=1$): $\gamma_0 = 0.72 \pm 0.08$ and  $\gamma_1 = 0.16^pm 0.34$. In these two last panels, the 2$\sigma$ interval corresponds to ($\Delta \chi^2=4.0$). The dashed areas show the results from simulations: $\gamma_0=0.81 \pm 0.06$ and $−0.02 \leq \gamma_1 \leq 0.07$ (Planelles et al. 2013). As one may see, the $\gamma_0$ value is in full agreement with cosmological hydrodynamical simulations. Finally,  Fig.(3d) shows the $\gamma(z)$ evolution (obtained by using the  1$\sigma$  results  from Figs.3b and 3c)  and the results from hydrodynamical simulations (dashed area). As one may see for this case,  the observational results are in full agreement with cosmological hydrodynamical simulations.

\subsection{Results by using $f_{gas}$ and $D_A$ from a flat $\Lambda$CDM model from the Planck results}

We also perform our method by using the 38 gas mass fraction measurements and  angular diameter distances  from the flat $\Lambda$CDM model constrained  by the current CMB observations (Ade et al. 2015). In this case,  the angular diameter distance for each galaxy cluster is given by Eq.(13) with $H_0$ and $\Omega_M$ as in Table I.

Our statistical results are plotted in fig.(4). Fig.(4a) shows the regions for 1$\sigma$  ($\Delta \chi^2=2.30$) and 2$\sigma$  ($\Delta \chi^2=6.17$)  on the $(\gamma_0,\gamma_1)$ plane for two free parameters. For this case, we obtain at 1$\sigma$: $\gamma_0=0.84 \pm 0.07$ and  $\gamma_1=-0.02 \pm 0.14$. Fig.(4b) shows the likelihood for $\gamma_0$ by marginalizing on $\gamma_1$ and Fig.(4c) shows the likelihood for $\gamma_1$ by marginalizing on $\gamma_0$. We obtain at 1$\sigma$ ($\Delta \chi^2=1$): $\gamma_0 = 0.84 \pm 0.05$ and  $\gamma_1 = -0.02 \pm 0.11$. The dashed areas show the results from simulations: $\gamma_0=0.81 \pm 0.06$ and $−0.02 \leq \gamma_1 \leq 0.07$ (Planelles et al. 2013). Finally,  Fig.(4d) shows the $\gamma(z)$ evolution (obtained by using the  1$\sigma$  results  from Figs.3b and 3c)  and the results from hydrodynamical simulations (dashed area). As one may see for this case,  the $\gamma(z)$ function shows  a negligible time evolution for gas mass fraction in galaxy clusters and the $\gamma_0$ and $\gamma_1$ values are in full agreement with cosmological hydrodynamical simulations.

\section{Discussion}

{  As commented earlier, we find a non-negligible time evolution to $\gamma(z)$  when the $D_A$ for the galaxy clusters in the La Roque et al. (2006) sample are those obtained by Bonamente et al. (2006). This fact may be due to spherical assumption used by these authors to describe the galaxy clusters and measure their $D_A$ via SZE plus X-ray observations. The importance of the intrinsic geometry of the cluster has been emphasized by many authors (Fox \& Pen 2002; Jing \& Suto 2002; Plionis, Basilakos \& Ragone-Figueroa 2006; Sereno et al. 2006) and the standard spherical geometry has been severely questioned, since {\it Chandra}  observations have shown that clusters usually exhibit an elliptical surface brightness. Moreover, by using $D_A$ of galaxy cluster samples, Holanda, Lima \& Ribeiro (2010,2011) and Meng et al. (2012) showed that that the ellipsoidal model is a better geometrical hypothesis describing the structure of the galaxy cluster compared with the spherical model if the cosmic distance duality relation, $D_L=D_A(1+z)^2$, is valid in cosmological observations.  As discussed by De Filippis et al. (2005), the choice of circular rather than elliptical model  affects significantly the values for core radius ($\theta_c$) and, consequently, the angular diameter distance, since $D_A$ $\alpha$ $\theta_c^{-1}$ (see also Holanda \& Alcaniz 2015 for a study about the elongation of the gas distribution of galaxy clusters based on measurements of the cosmic expansion rate, luminosity distance to type Ia supernovae and angular diameter distances of galaxy clusters).

  On the other hand, our results for $\gamma(z)$ by considering angular diameter distances for the galaxy clusters in the La Roque et al. (2006) sample from Planck results (Ade et al. 2015) and from a  sample of galaxy clusters described by an elliptical profile (De Filippis et al. 2005) show that  the non-isothermal double $\beta$-model does not affect the quality of gas mass fraction measurements in La Roque et al. sample (2006) if one considers that a constant $\gamma(z)$ is an inherent characteristic  these structures. However, it affects their angular diameter distances. }

\section{Conclusions}

The study of gas mass fraction in galaxy clusters as well its apparent evolution with redshift can be used to constrain cosmological parameters.  An important quantity in this context is the depletion factor, $\gamma(z)$, the ratio by which the measured baryon fraction  in galaxy clusters is depleted with respect to the universal.  Cosmological hydrodynamical simulations where a specific flat $\Lambda$CDM model is considered have been used to determinate it. Very recently, Holanda et al. (2017) proposed the first self-consistent observational constraints on the gas depletion factor by using SNe Ia and X-ray gas mass fractions observations. The sample of $f_{gas}$ measurements in their analyses were taken on a  $(0.8-1.2)$ $\times r_{2500}$ shell (Mantz et al. 2014). No time evolution for $\gamma(z)$ was obtained.

In this paper, by using exclusively  galaxy cluster data, we proposed a new method to obtain $\gamma (z)$ that uses two data set of the same galaxy clusters. We considered 38 $f_{gas}$ and angular diameter distance measurements from galaxy clusters where the non-isothermal spherical  double $\beta$  model was used to describe them (La Roque et al. 2006; Bonamente et al. 2006). A possible $\gamma(z)$ time evolution was parametrized by $\gamma(z) = \gamma_0(1 + \gamma_1z)$. Our results showed that: the $\gamma_0$ value is in full agreement with the most recent cosmological hydrodynamical simulations (Planelles et al. 2013; Battaglia et al. 2013), however, we obtain $\gamma_1=-0.42^{+0.27}_{-0.23}$ at 1$\sigma$, indicating a  time evolution for the baryon depletion factor (see Fig.2c). Since some galaxy clusters presented questionable  reduced $\chi^{2}$ ($2.43 \leq \chi_{d.o.f.}^2 \leq 41.62$) when described by the non-isothermal double $\beta$ model (see Table 6 in Ref. Bonamente et al. 2006), we excluded these data and the results still showed a non-negligible time evolution for $\gamma(z)$ (see dashed lines in Fig.2b and 2c).

{  However, we also considered in our analyses angular diameter distances for the galaxy clusters in the La Roque et al. (2006) sample  from the flat $\Lambda$CDM model constrained by the current CMB observations (Ade et al. 2015) and from an angular diameter distance sample of galaxy clusters described by an elliptical profile (De Filippis et al. 2005). For these cases, a negligible time evolution for gas mass fraction in galaxy clusters was obtained and the $\gamma_0$ and $\gamma_1$ values  are in full agreement with cosmological hydrodynamical simulations (see Figs. 3b, 3c, 4b and 4c). Then, the  tension between the non-isothermal double $\beta$-model and a constant $\gamma(z)$ function in La Roque et al. (2006) sample found in our analyses may be due to spherical assumption used by Bonamente et al. (2006) to describe the galaxy clusters. The choice of circular rather than elliptical model  affects significantly the values for core radius ($\theta_c$) and, consequently, the angular diameter distance ($D_A$ $\alpha$ $\theta_c^{-1}$) (See de Filippis et al. 2005). Therefore, as a general conclusion, if a constant $\gamma(z)$ is an inherent characteristic of these structures, our results  showed that using the non-isothermal double $\beta$-model did not affect the quality of the gas mass fraction measurements in La Roque et al. Sample (2006), but it affected their angular diameter distances.}

\section*{Acknowledgments}
RFLH acknowledges financial support from INCT-A and CNPq (No. 478524/2013-7, 303734/2014-0).

\end{document}